\documentclass[twocolumn,amsmath,amssymb,prl,showpacs]{revtex4}
\usepackage{graphicx}

\def\d{{\partial}}
\def\s{{\sigma}}
\def\e{{\epsilon}}
\def\k{{ {\bf k} }}

\def\bfl{{ {\mbox{\boldmath $l$}} }}
\def\bfs{{ {\mbox{\boldmath $s$}} }}
\def\w{{\omega}}
\def\a{{\alpha}}
\def\b{{\beta}}

\def\g{{\gamma}}
\def\l{{\lambda}}

\begin{document}

\def\runtitle{
Study of intrinsic spin and orbital Hall effects in Pt \\
}
\def\runauthor{
H. {\sc Kontani}$^1$, M. {\sc Naito}$^1$, D.S. {\sc Hirashima}$^1$, 
K. {\sc Yamada}$^2$ and J. {\sc Inoue}$^3$
}

%\draft
\title{
Study of intrinsic spin and orbital Hall effects in Pt \\
based on a ($6s, 6p, 5d$) tight-binding model
}

\author{
H. {\sc Kontani}$^1$, M. {\sc Naito}$^1$, D.S. {\sc Hirashima}$^1$, 
K. {\sc Yamada}$^2$ and J. {\sc Inoue}$^3$
}

\address{
$^1$Department of Physics, Nagoya University,
Furo-cho, Nagoya 464-8602, Japan. \\
$^2$College of Science and Engineering, Ritsumeikan University, 
%1-1-1 Noji Higashi, 
Kusatsu, Shiga 525-8577, Japan. \\
$^3$Department of Applied Physics, Nagoya University,
Furo-cho, Nagoya 464-8602, Japan. 
}

\date{\today}

\begin{abstract}
We study the origin of the intrinsic spin Hall conductivity (SHC) 
and the $d$-orbital Hall conductivity (OHC) in Pt based on a 
multiorbital tight-binding model with spin-orbit interaction.
We find that the SHC exceeds
$1000\ \hbar e^{-1}\!\cdot\!\Omega^{-1}{\rm cm}^{-1}$
when the resistivity $\rho$ is smaller than $\sim10 \ \mu\Omega$ cm,
whereas it decreases to
$300 \ \hbar e^{-1}\!\cdot\!\Omega^{-1}{\rm cm}^{-1}$
when $\rho\sim 100\ \mu\Omega$ cm.
In addition, the OHC is still larger than the SHC.
The origin of the huge SHE and OHE in Pt is the large 
``effective magnetic flux''
that is induced by the interorbital transition between $d_{xy}$- and 
$d_{x^2-y^2}$-orbitals with the aid of the strong spin-orbit interaction.
\end{abstract}


\sloppy

\maketitle

%%%%%%%%%%%%%%%%%%
% Introduction
%%%%%%%%%%%%%%%%%%

Recently, the spin Hall effect (SHE) has attracted much attention due to 
its fundamental interest and its potential application in spintronics.
The SHE has a close relation to the anomalous 
Hall effect (AHE) in ferromagnets:
In 1954, Karplus and Luttinger (KL) \cite{Karplus} studied the 
Hall effect in multiband systems and found that an electric field 
induces a spin-dependent transverse current in the presence of 
spin-orbit (SO) interaction.
This effect causes the AHE (transverse charge current) 
in ferromagnetic metals and the SHE (transverse spin current)
in paramagnetic metals.
These phenomena are fundamental issues in recent condensed matter physics
 \cite{Murakami,Niu04,inoue1,raimondi,rashba,inoue2,kato,wunderlich,valenzuela,ZnSe,Pt-Saitoh,kimura,Kontani07,Kontani06,Miyazawa,Kontani94}.
In these years, great progress on the SHE in semiconductors has been made.
Murakami et al. \cite{Murakami} and Sinova et al. \cite{Niu04}
have studied the intrinsic (impurity-independent) SHE in semiconductors 
by developing the theory of KL.
Now, the SHE in two-dimensional electron gas (2DEG) with a Rashba-type 
SO interaction is well understood \cite{inoue1,raimondi,rashba,inoue2}.
Although the SHE in semiconductors was recognized by the optical detection 
of spin accumulation \cite{kato,wunderlich},
it is unfortunately too small for quantitative analysis.
Therefore, materials that show a large SHE are highly desirable.

Recent experiments have revealed that the SHE also exists
in metals such as Al \cite{valenzuela} and Cl-doped ZnSe \cite{ZnSe}.
In particular, the huge spin Hall conductivity (SHC) in Pt at room temperature
[$240\ \hbar e^{-1} \cdot \Omega^{-1}{\rm cm}^{-1}$] \cite{kimura},
which is $10^4$ times larger than the SHC reported in semiconductors,
has attracted great attention.
%in connection with possible technological applications.
Simple 2DEG models cannot explain this experimental fact.
Recently, the present authors have studied the SHE in Sr$_2$RuO$_4$ 
that is described by the $t_{2g}$-orbital tight-binding model
\cite{Kontani07} and found that the anomalous velocity due to 
interorbital hopping gives rise to huge SHC in transition metals.
%It is the first theoretical study of SHE in transition metal complexes.
This mechanism also causes the large AHE \cite{Kontani06,Miyazawa,Kontani94}.
To reveal the origin of the huge SHE in Pt, we have to investigate
the anomalous velocity due to the multiorbital effect
by considering all the $d$-orbitals ($t_{2g}$+$e_{g}$ orbitals).
%Because of the generality of $t_{2g}$-model,
%giant SHE should be observed universally in multiorbital $d$-electron systems.

In this letter, we study the intrinsic SHE and the 
$d$-orbital Hall effect (OHE) in Pt by analyzing a realistic
multiorbital tight-binding model.
In the low-resistivity regime where $\rho <10 \ \mu\Omega$cm,
both the SHC and orbital Hall conductivity (OHC) are constant
of order $1000\sim3000$ $\hbar e^{-1} \cdot \Omega^{-1}{\rm cm}^{-1}$,
whereas they are strongly suppressed in the high-resistivity regime
where $\rho \gg 10 \ \mu\Omega$cm.
The derived coherent-incoherent crossover
is a universal property of intrinsic Hall effects \cite{Kontani94,Kontani06}.
Both the SHE and OHE originate from a kind of Peierls phase factor 
due to the ``effective magnetic flux'' \cite{Kontani07}
that is induced by a combination of the angular dependence 
of $d$-orbital wave functions and SO interaction.
In Pt, the dominant contribution to the SHE is given by
the $d_{xy}$-orbital (in $t_{2g}$) 
and the $d_{x^2-y^2}$-orbital (in $e_g$).
Therefore, both the $t_{2g}$- and $e_g$-orbitals should be taken 
into account to explain the huge SHE in Pt.

%%%%%%%%%%%%%%%%%%%%%%%%%%%%%%%%%%%%%%%%%%%%%%%%%%%%%%
\begin{figure}[htbp]
\begin{center}
\includegraphics[width=.7\linewidth]{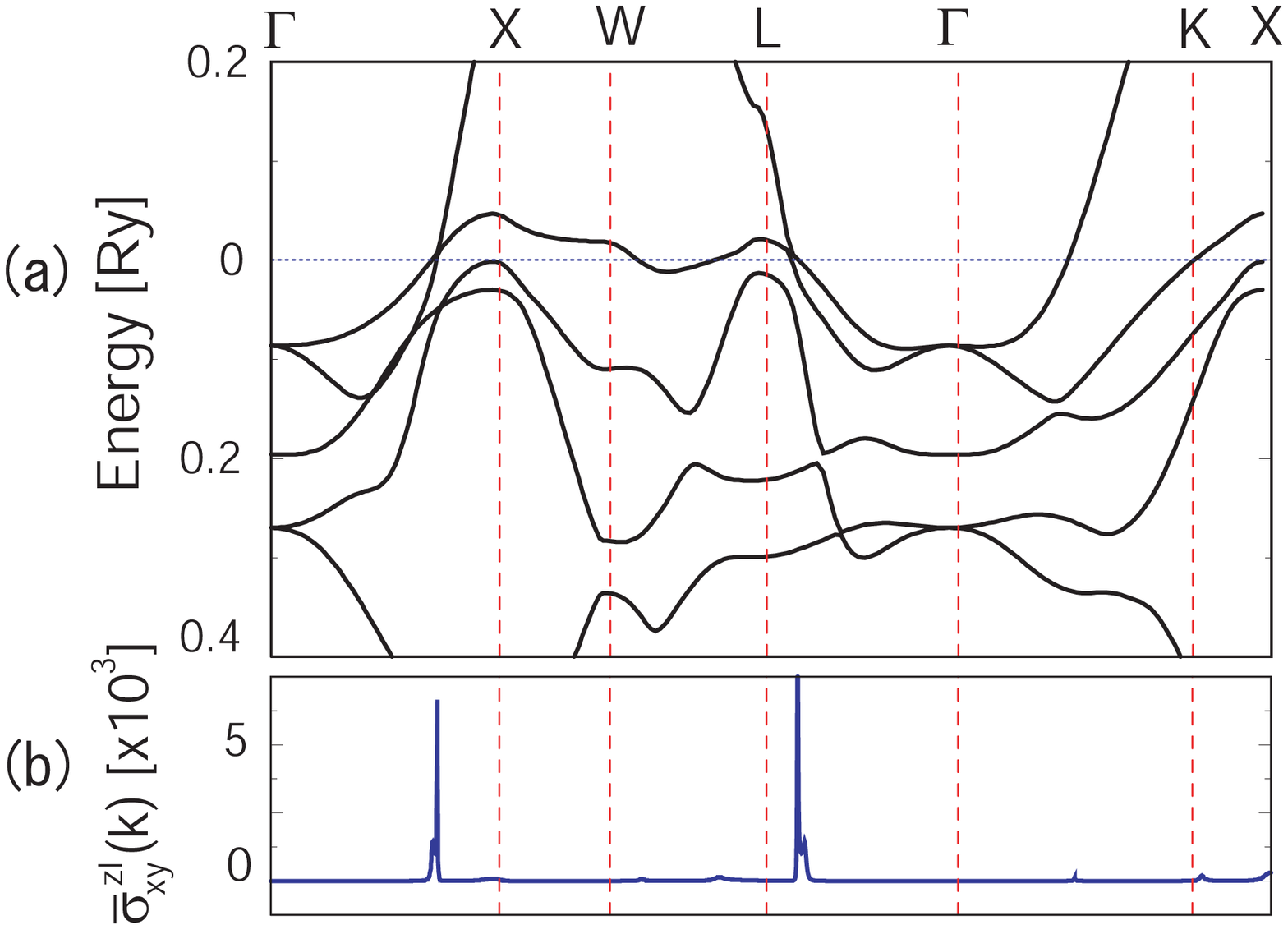}\\
\includegraphics[width=.75\linewidth]{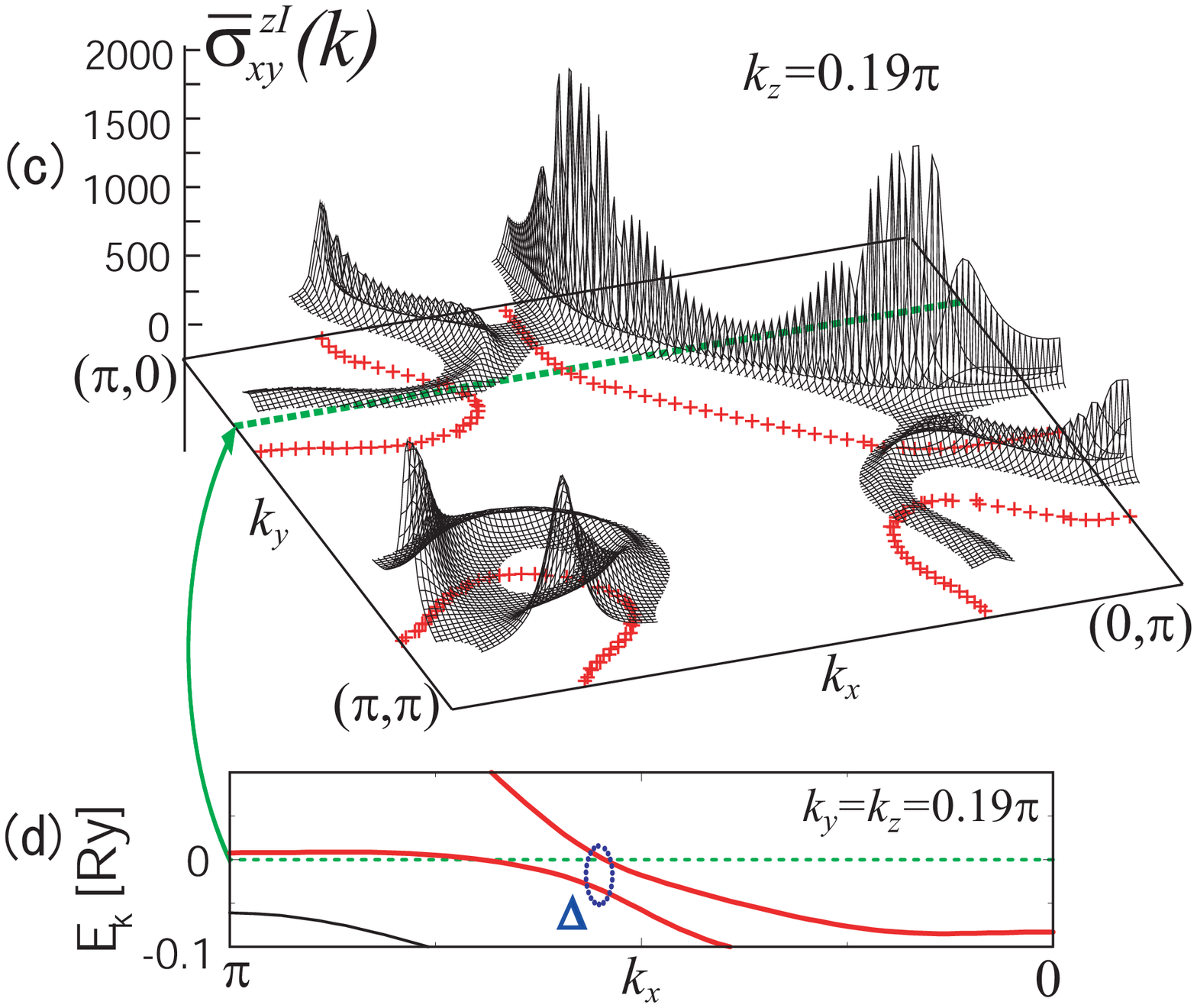}
\end{center}
\caption{(Color online) 
(a) Band structure of a ($6s, 6p, 5d$) tight-binding model
for Pt ($\l=0.04$).
$\Gamma \ = \ (0,0,0)$, X$\ = \ (\pi,0,0)$, W$\ = \ (\pi,\pi/2,0)$,
L$\ = \ (\pi/2,\pi/2,\pi/2)$, and K$\ = \ (3\pi/4,3\pi/4,0)$.
(b) ${\bar \s}_{xy}^z(\k)$ for $\gamma=0.002$ along $\Gamma$-X.
(c) ${\bar \s}_{xy}^z(\k)$ on the plane $k_z=0.19\pi$, 
which is plotted only for $\s_{xy}^z(\k)>25$.
The crosses represent the Fermi surfaces.
(d) $k_x$-dependence of $E_\k^l$ for $k_y=k_z=0.19\pi$; 
the minimum bandsplitting near the Fermi level $\Delta$ is $\sim 0.035$.
}
  \label{fig:band}
\end{figure}
%%%%%%%%%%%%%%%%%%%%%%%%%%%%%%%%%%%%%%%%%%%%%%%%%%%%%

Pt has a face-centered cubic (FCC) structure with $a=3.9$ \AA, 
and the nearest inter-atomic distance is 2.8 \AA.
% Hirashima
In the present study,
we use the Naval Research Laboratory tight-binding (NRL-TB) 
model \cite{NRL1,NRL2} to
describe the bandstructure in Pt. The NRL-TB model employs the scheme
of the two-center, non-orthogonal Slater-Koster (SK) Hamiltonian \cite{SK}.
The SK parameters are represented with distance- and environment-dependent 
parameters that are determined so that the obtained total
energy and the band structures agree well with those obtained by the
first-principles calculations.
%To describe the electronic state in Pt,
Here, we take into account $6s$, $6p$, and $5d$ orbitals (in total, nine) 
and hopping integrals up to the sixth nearest neighbor sites.
The electron number per Pt atom is ten.
The NRL-TB model uses non-orthogonal bases, but we neglect the overlap
integrals between different sites in this study for simplicity \cite{comment}. 
This simplification significantly changes the dispersion of the $s$-band 
far from the Fermi level, while that of the band structure near the 
Fermi energy is little affected.
In the presence of the SO interaction for $5d$ electrons 
$H_{\rm SO}=\lambda \sum_{i} (\bfl\!\cdot\!\bfs)_i$,
the total Hamiltonian becomes
\begin{eqnarray}
{\hat H}=\left(
\begin{array}{cc}
 {\hat H}_0 +\l {\hat l}_z/2 & \l({\hat l}_x-i{\hat l}_y)/2 \\
 \l({\hat l}_x+i{\hat l}_y)/2 & {\hat H}_0 -\l {\hat l}_z/2 
\end{array}
\right) ,
\end{eqnarray}
where ${\hat H}_0$ is a $9\times9$ matrix given by the NRL-TB model.
The matrix elements of $\bfl$ are given in ref. \cite{Friedel}.
The bandstructure obtained for the ($6s, 6p, 5d$) tight-binding model
with $\l=0.04$ Ry is shown in Fig. \ref{fig:band} (a),
which is in good agreement with the result of a relativistic 
first-principles calculation \cite{And,Bei} near the Fermi level.
According to optical spectroscopy, $\l=0.03$ Ry for a $5d$ electron in Pt,
and  $\l=0.013$ Ry for a $4d$ electron in Pd \cite{Friedel}.
Hereafter, we set the unit of energy Ry; 1 Ry = 13.6 eV.
Based on the NRL-TB model, spin wave excitations and the electron 
self-energy corrections in the ferromagnetic Fe are studied 
using the random-phase approximation \cite{Naito-Fe}.

The $18\times18$ matrix form of the
retarded Green function is given by
${\hat G}^R({\k},\w)=(\w+\mu-{\hat H}+i{\hat \Gamma})^{-1}$,
where $\mu$ is the chemical potential 
and ${\hat \Gamma}$ is the imaginary part of the $\k$-independent
self-energy (damping rate) due to scattering by local impurities
(or inelastic scattering by phonons).

The charge current in the present model is
\begin{eqnarray}
{\hat J}_\mu^{\rm C} &=& \left(
\begin{array}{cc}
 {\hat j}_\mu^{\rm C} & 0 \\
 0 & {\hat j}_\mu^{\rm C} \\
\end{array}
\right) \label{eqn:J}.
\end{eqnarray}
Here, ${\hat j}_{\mu}^{\rm C} = -e \frac{\d {\hat H}_0}{\d k_\mu}$,
where $-e$ is the electron charge and $\mu=x,y$.
In this case, the atomic SO interaction is not involved in the charge current
since it is $\k$-independent.
Then, the $s_z$-spin current
${\hat J}_\mu^{\rm S}= \{ {\hat J}_\mu^{\rm C}, {\hat s}_z \}/2$ 
is expressed as
\begin{eqnarray}
{\hat J}_\mu^{\rm S} &=& (-\hbar/e)\left(
\begin{array}{cc}
 {\hat j}_\mu^{\rm C} & 0 \\
 0 & -{\hat j}_\mu^{\rm C} \\
\end{array}
\right) \label{eqn:JS}.
\end{eqnarray}
%
%Since $\langle {\hat J}_\mu^{\rm C,S}\rangle_{\rm FS}=0$, 
%current vertex correction due to short-ranged impurities is absent 
%\cite{Kontani06,Kontani07}.

Here, we discuss the current vertex correction (CVC) due to the 
local impurity potentials in the Born approximation, which is given by
$\Delta{\hat J}_\mu^C \propto \sum_\k {\hat G}^A {\hat J}_\mu^C {\hat G}^R$.
When $\a$ is one of the $p$-orbitals and $\b$ is one of the $(s,d)$-orbitals,
$({\hat H}_0(\k))_{\a,\b}$ is an odd function 
with respect to $\k \leftrightarrow -\k$, and therefore
the $(\a,\b)$-component of $(\d/\d k_\mu){\hat G} = 
{\hat G} {\hat J}_\mu^C {\hat G}$ is an even function.
Note that $|p_\nu\rangle \rightarrow -|p_\nu\rangle$ ($\nu=x,y,z$)
under the parity transformation.
Thus, $(\Delta{\hat J}_\mu^C)_{\a,\b}$ is finite
only when either $\a$ or $\b$ is a $p$-orbital.
In Pt, however, we have verified that the CVC affects the SHE only 
slightly [less than 5\%] since the $6p$-level is 20 eV higher than 
the Fermi level $\mu$ and the $p$-electron density of states (DOS) 
at $\mu$ is very small. 
For this reason, we disregard the CVC hereafter.

According to the linear response theory \cite{Streda}, 
the SHC is given by $\s_{xy}^z = \s_{xy}^{zI}+\s_{xy}^{zI\!I}$, where
\begin{eqnarray}
\s_{xy}^{zI} &=& \frac{1}{2\pi N}\sum_{\k}
{\rm Tr}\left[{\hat J}_x^{\rm S} {\hat G}^R {\hat J}_y^{\rm C} {\hat G}^A 
\right]_{\w=0},  \label{eqn:SHCI}
 \\
\s_{xy}^{zI\!I} &=& \frac{-1}{4\pi N} \sum_{\k}\int_{-\infty}^0 d\w
{\rm Tr}\left[{\hat J}_x^{\rm S} \frac{{\hat G}^R}{\d\w} 
{\hat J}_y^{\rm C} {\hat G}^R \right.
 \nonumber \\
& &\left. \ \ \ \ \ \ \ \ 
 -{\hat J}_x^{\rm S} {\hat G}^R
{\hat J}_y^{\rm C} \frac{{\hat G}^R}{\d\w} 
- \langle {\rm R}\rightarrow {\rm A} \rangle
\right].
 \label{eqn:SHCII}
\end{eqnarray}
Here, $I$ and $I\!I$ represent the ``Fermi surface term''
and the ``Fermi sea term'', respectively.
In the same way, the OHC of the Fermi surface term $O_{xy}^{zI}$ 
and that of the Fermi sea term $O_{xy}^{zI\!I}$
are given by eqs. (\ref{eqn:SHCI}) and (\ref{eqn:SHCII}), respectively,
by replacing ${\hat J}_x^{\rm S}$ with the $l_z$-orbital current
${\hat J}_x^{\rm O}= \{ {\hat J}_x^{\rm C}, {\hat l}_z \}/2$.
%Note that $\s_{xy}^{z}=\s_{yz}^{x}=\s_{zx}^{y}$ and
%$O_{xy}^{z}=O_{yz}^{x}=O_{zx}^{y}$ because of the cubic symmetry of Pt.
Because of the cubic symmetry of Pt,
$\s_{\mu\nu}^{\delta}=\s_{xy}^{z}\cdot\e_{\mu\nu\delta}$ and
$O_{\mu\nu}^{\delta}=O_{xy}^{z}\cdot\e_{\mu\nu\delta}$, where 
$\mu,\nu,\delta=x,y,z$ and $\e_{\mu\nu\delta}$ is the antisymmetrized 
tensor with $\e_{xyz}=1$.

When $\Gamma_{\a\b}=\gamma\delta_{\a\b}$ (constant $\gamma$ approximation), 
$\w$-integration in eq. (\ref{eqn:SHCII}) can be performed analytically 
as shown in ref. \cite{Kontani06}:
Then, $\s_{xy}^{zI\!I}= \s_{xy}^{zI\!Ia}+\s_{xy}^{zI\!Ib}$, where 
\begin{eqnarray}
\s_{xy}^{zI\!Ia}&=&
\frac{-1}{2\pi N} \sum_{\k,l\ne m} 
 {\rm Im}\left\{ (J_x^{\rm S})^{m l}(J_y^{\rm C})^{l m} \right\}
 \frac1{E_\k^l -E_\k^m}
 \nonumber \\
& &\times {\rm Im} \left\{ \frac{E_\k^l +E_\k^m -2i\gamma}
 {(E_\k^l -i\gamma)(E_\k^m -i\gamma)} \right\} ,
 \label{eqn:AP-II1} \\
\s_{xy}^{zI\!Ib}&=&
\frac{1}{\pi N} \sum_{\k,l\ne m} 
 {\rm Im}\left\{ (J_x^{\rm S})^{m l}(J_y^{\rm C})^{l m} \right\}
 \frac1{(E_\k^l -E_\k^m)^2} 
 \nonumber \\
& &\times {\rm Im} \left\{ 
 \ln \left( \frac{E_\k^l -i\gamma}{E_\k^m -i\gamma}
 \right) \right\} ,
 \label{eqn:AP-II2}
\end{eqnarray}
where $l,m$ represent the band indices.
$E_\k^l$ is the $l$th eigenenergy of ${\hat H}$ measured from the 
chemical potential $\mu$;
$\sum_{\a\b}U_{l\a}^\dagger H_0^{\a\b} U_{\b m}= E_\k^l\delta_{lm}$,
where $\a,\b$ are the orbital indices and 
$U$ is a $\k$-dependent unitary matrix.
$(J_x^{\rm S})^{m l}$ in eqs. (\ref{eqn:AP-II1}) and 
(\ref{eqn:AP-II2}) is given by $\sum_{\a\b}U_{l\a}^\dagger 
(J_x^{\rm S})^{\a\b} U_{\b m}$.

In the Born approximation, ${\hat \Gamma}$ in the Green function is given by 
$n_{\rm imp}I^2\frac1N\sum_\k{\rm Im}({\hat G}^A(0)-{\hat G}^R(0))$,
where $I$ is the local impurity potential and $n_{\rm imp}$ is the 
impurity density.
When $\lambda/E_{\rm F}\ll 1$, ${\hat \Gamma}$ is almost diagonal 
with respect to the orbital index; $\Gamma_{\a,\b}=\g_{\a}\delta_{\a,\b}$
 \cite{Kontani07}.
In the case of Sr$_2$RuO$_4$, the SHC %in the clean limit
in the Born approximation is nearly three times greater than that 
in the constant $\g$ approximation \cite{Kontani07}.
In Pt, in contrast, we have verified that both approximations give a
similar SHC in the clean limit.
For this reason, we use the constant $\gamma$ approximation hereafter.

Here, we determine the part of the Fermi surface from which the SHC originates:
In Fig. \ref{fig:band} (b) and (c), we show
$\displaystyle {\bar \s}_{xy}^{zI}(\k) \equiv \frac18 \sum_{k_x'= \pm k_x} 
\sum_{k_y'= \pm k_y} \sum_{k_z'= \pm k_z} \s_{xy}^{zI}(\k')$,
where $\s_{xy}^{zI}(\k)$ is the integrand in eq. (\ref{eqn:SHCI}).
%$\displaystyle
%\s_{xy}^z(\k)\equiv \frac18 
%\sum_{(k_x',k_y',k_z')}^{(\pm k_x,\pm k_y, \pm k_z)}
%{\rm Tr}\left[{\hat J}_x^{\rm S} 
%{\hat G}^R {\hat J}_y^{\rm C} {\hat G}^A \right]_{\k',\w=0}.
%$
[Apparently, $\frac{1}{N}\sum_{\k}{\bar \s}_{xy}^{zI}(\k)= \s_{xy}^{zI}$.] 
${\bar \s}_{xy}^z(\k)$ is finite only on the Fermi surface, 
and it takes huge values at $(0.73\pi,0,0)$ (on $\Gamma$-X)
and at $(0.42\pi,0.42\pi,0.42\pi)$ (on L-$\Gamma$) since 
two bands are very close on the Fermi level in the present model.
However, the contribution of these two points to the SHC
is small after taking the $\k$-summation.
The dominant contribution comes from a wide area around
$(0.19\pi,0.19\pi,0.57\pi)$ as shown in Fig. \ref{fig:band} (d).
Here, the bandsplitting $\Delta$ near the Fermi level is $0.035$.

%%%%%%%%%%%%%%%%%%%%%%%%%%%%%%%%%%%%%%%%%%%%%%%%%%%%%%
\begin{figure}[htbp]
\begin{center}
\includegraphics[width=.75\linewidth]{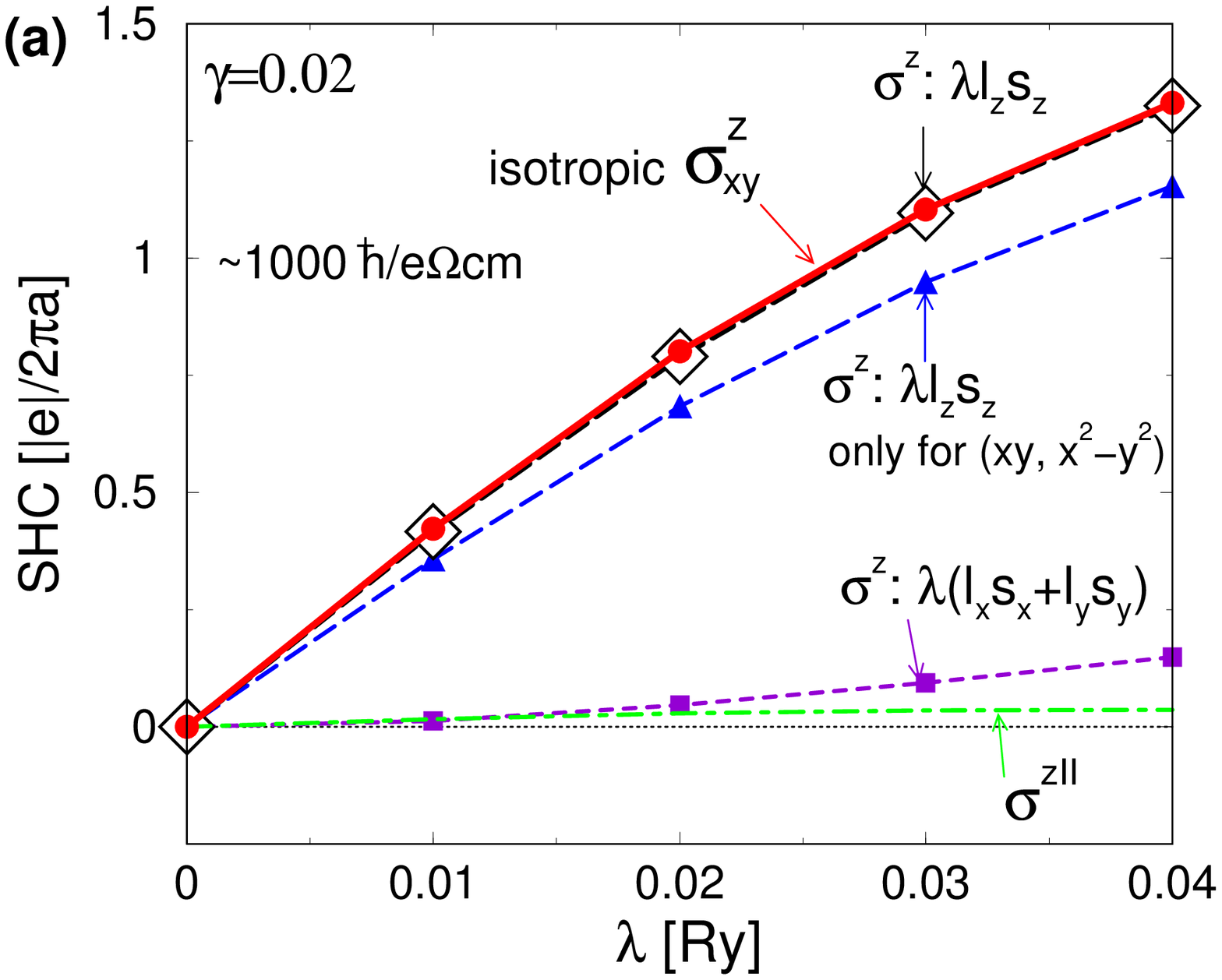}
\includegraphics[width=.8\linewidth]{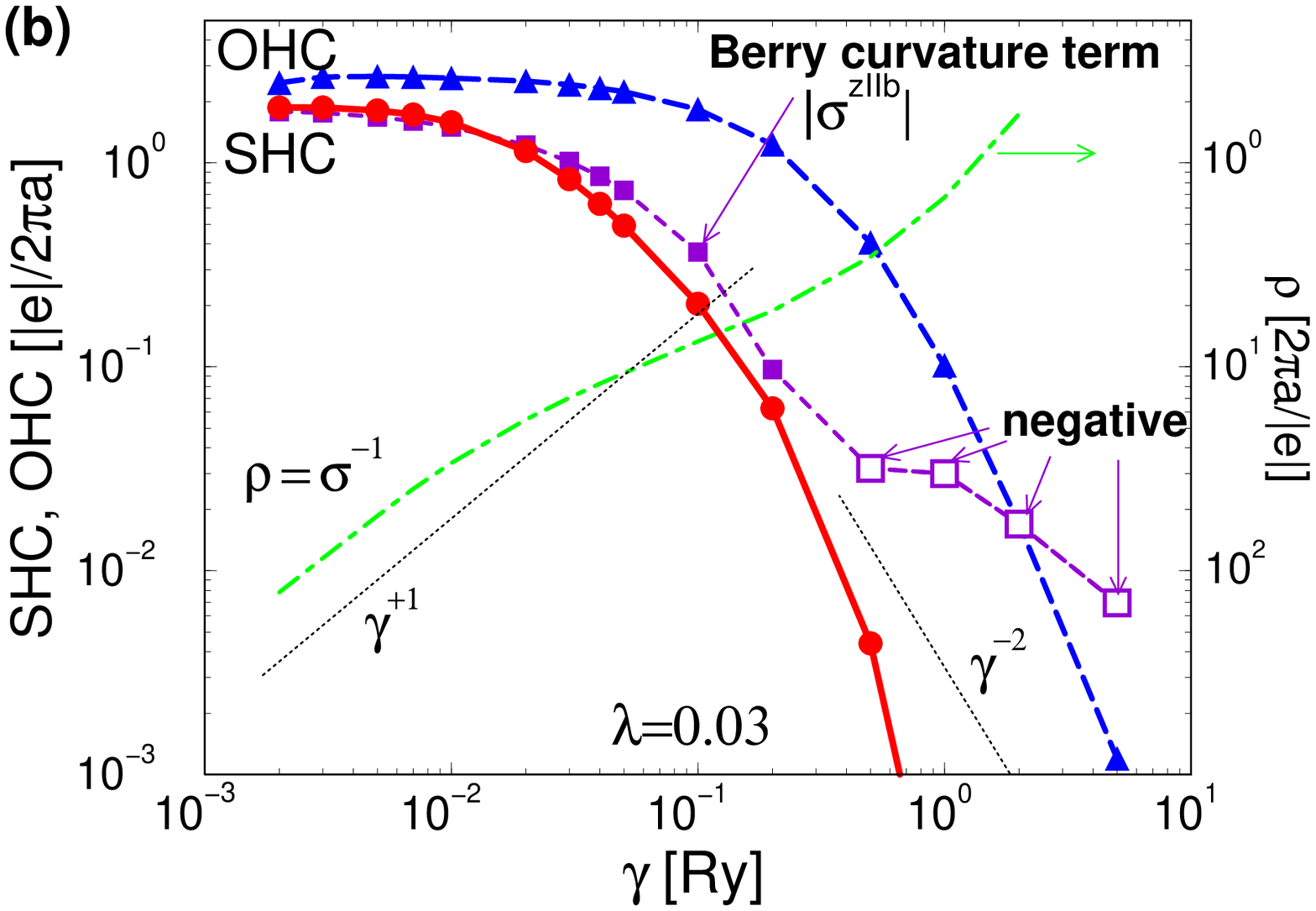}
\end{center}
\caption{(Color online)
(a) $\l$-dependence of the SHC.
The relation $\s_{xy}^z \approx \s_{xy}^{zI} \gg \s_{xy}^{zI\!I}$ is realized.
The matrix element of the SO interaction between the $d_{xy}$-orbital and 
the $d_{x^2\mbox{-}y^2}$-orbital gives the dominant contribution to the SHC.
Note that $1 \ [|e|/2\pi a] \approx 
1000 \ \hbar e^{-1}\cdot\Omega^{-1}{\rm cm}^{-1}$ for $a=4$ \AA.
(b) Crossover behaviors of the SHC and OHC at $\g \sim\Delta \approx 0.035$.
$\rho$ is approximately proportional to $\g$.
$\rho=1$ corresponds to $1000 \ \mu\Omega{\rm cm}$.
}
  \label{fig:SPEC}
\end{figure}
%%%%%%%%%%%%%%%%%%%%%%%%%%%%%%%%%%%%%%%%%%%%%%%%%%%%%

Now, we perform the numerical calculation of the SHC,
using $128^3 \sim 512^3$ $\k$-meshes.
Figure \ref{fig:SPEC} (a) shows the $\l$-dependence of the total SHC
$\s_{xy}^{z}=\s_{xy}^{zI}+\s_{xy}^{zI\!I}$ for $\g=0.02$,
%when $\Gamma_{\a\b}=\gamma\delta_{\a\b}$ with $\g=0.02$, 
which is smaller than $\Delta=0.035$.
$\s^{zI\!I}$ represents the Fermi sea term in eq. (\ref{eqn:SHCII}).
Apparently, $\s_{xy}^{z} \approx \s_{xy}^{zI}\gg \s_{xy}^{zI\!I}$ is realized.
$\s_{xy}^z$ increases with $\l$ monotonically, and it reaches
$1000 \ \hbar e^{-1}\cdot\Omega^{-1}{\rm cm}^{-1}$ at $\l=0.03$.
To clarify the origin of the SHE,
we study the SHC when the SO interaction is anisotropic:
As shown in Fig. \ref{fig:SPEC} (a),
$\s_{xy}^z$ for $H_{\rm SO}=\l\sum_i(l_x s_x + l_y s_y)_i$ is much smaller than
that in the isotropic case where $H_{\rm SO}=\l\sum_i(\bfl\!\cdot\!\bfs)_i$].
On the other hand, $\s_{xy}^z$ for $H_{\rm SO}=\l\sum_i(l_z s_z)_i$ 
almost coincides with that in the isotropic case.
Therefore, it is concluded that the $z$-component of the SO interaction
gives the decisive contribution to the SHC.
The matrix element of $l_z$ is finite only for
$\langle yz| l_z |zx\rangle=-\langle zx| l_z |yz\rangle=i$ and 
$\langle xy| l_z |x^2\mbox{-}y^2\rangle=
-\langle x^2\mbox{-}y^2| l_z |xy\rangle=2i$.
Among them, the $d_{xy}$- and $d_{x^2\mbox{-}y^2}$-orbitals,
both of which are given by the linear combinations of $l_z=\pm2$, 
cause a dominant contribution to the SHC as shown in Fig. \ref{fig:SPEC} (a).

Here, we discuss the $\gamma$-dependence of the SHC and OHC:
When $\gamma$ is sufficiently small, these intrinsic Hall conductivities
%are independent of $\gamma$ 
%since they are caused by interband particle-hole excitation 
are proportional to the lifetime of 
the interband particle-hole excitation: $\hbar/\Delta$
 \cite{Karplus,Kontani94,Murakami,Niu04,Kontani06,Kontani07}.
In fact, Fig. \ref{fig:SPEC} (b) shows that both the SHC and OHC 
for $\l=0.03$ \cite{Friedel}
are independent of $\g$ for $\g \ll\Delta\sim0.035$.
In the high-resistivity regime where $\g \gg \Delta$,
both SHC and OHC decrease drastically with $\gamma$ since the 
interband excitation is suppressed when the quasiparticle lifetime 
$\hbar/\g$ is shorter than $\hbar/\Delta$. 
This coherent-incoherent crossover of the intrinsic Hall conductivities
($\s_{xy}=$ const. for $\g\ll\Delta$ and 
$\s_{xy}\propto \rho^{-2}$ for $\g\gg\Delta$)
has been analyzed theoretically in refs. \cite{Kontani94,Kontani06}.
In Pt, the SHC decreases much faster than $\rho^{-2}$
in the high-resistivity regime and the SHC becomes negative for $\g>1$,
which may be due to a complex multiband structure.
If we put $\g\sim 0.07$, 
$\rho\sim 0.1\ [|e|/2\pi a]\sim100\ \mu\Omega{\rm cm}$. 
Then, the obtained SHC is
$\sim 300 \ \hbar e^{-1} \cdot \Omega^{-1}{\rm cm}^{-1}$,
which is close to the experimental SHC of Pt \cite{kimura}.
In the experimental situation, 
$\gamma$ in Fig. \ref{fig:SPEC} (b) corresponds to
$\hbar/2\tau$ within the spin diffusion length in Pt ($\sim$10 nm) 
from the interface of the junction, 
which might be larger than the bulk value of $\gamma$.

We comment on $\s_{xy}^{zI\!Ib}$, 
which is frequently called the ``Berry curvature term''.
When $\Gamma_{\a\b}=\gamma\delta_{\a\b}$ and $\gamma\rightarrow0$,
$\s_{xy}^z = \s_{xy}^{zI\!Ib} \approx \s_{xy}^{zI}$ \cite{Kontani06}.
However, $\s_{xy}^{zI\!Ib}$ is totally different from
$\s_{xy}^z$ in the high-resistivity regime as shown in Fig. \ref{fig:SPEC} (b),
since the cancellation between $\s_{xy}^{zI}$ and $\s_{xy}^{zI\!Ia}$ 
becomes worse when $\gamma$ is large.
In many systems including Pt, $\s_{xy}^z \approx \s_{xy}^{zI}$ is realized for 
a wide range of parameters  \cite{Kontani06}.

We briefly discuss the SHC using the Born approximation, where $\g_\a$ 
is proportional to the DOS for the $\a$-orbital, $\rho_\a(0)$.
When $\gamma_\a$ is $\a$-dependent, 
$\s_{xy}^{zI\!Ib} \ne \s_{xy}^z $ even in the clean limit \cite{Kontani07}.
In fact, the SHC in Sr$_2$RuO$_4$ given by the Born approximation
is much larger than that given by the constant $\g$ approximation
since the $\a$-dependence of $\rho_\a(0)$ is large \cite{Kontani07}.
In Pt, however, both approximations give similar results.
For this reason, we use the constant $\gamma$ approximation.

%%%%%%%%%%%%%%%%%%%%%%%%%%%%%%%%%%%%%%%%%%%%%%%%%%%%%%
\begin{figure}[htbp]
\includegraphics[width=.7\linewidth]{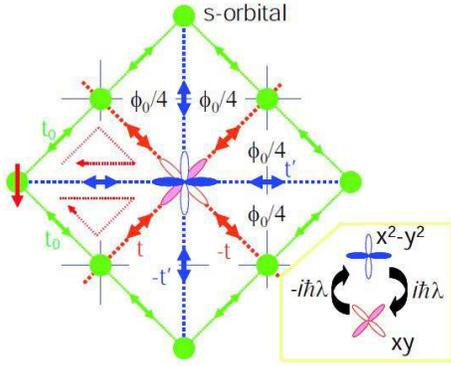}
\caption{(Color online)
Effective magnetic flux for $\uparrow$-electron
in the two-dimensional Pt model.
This is the origin of the huge SHC and AHC in Pt.
}
  \label{fig:flux}
\end{figure}
%%%%%%%%%%%%%%%%%%%%%%%%%%%%%%%%%%%%%%%%%%%%%%%%%%%%%

Figure \ref{fig:flux} shows the FCC crystal structure 
of Pt on the $xy$-plane.
Based on this two-dimensional model, we explain an intuitive reason 
why the huge SHC appears in Pt, by considering only $d_{xy}$-, 
$d_{x^2\mbox{-}y^2}$-, and $s$-orbitals.
$\pm t$ represents the hopping integrals between
the nearest neighbor $d_{xy}$-orbital and $s$-orbital, and
$\pm t'$ is used for the next nearest neighbor
$d_{x^2\mbox{-}y^2}$-orbital and $s$-orbital.
Both the hopping integrals change their signs by rotation by $\pi/2$.
%Since the kinetic terms due to $t$ is $4t\sin k_x a \sin k_y a$,
%it gives the anomalous velocity $v_x^a = 4at\cos k_x a \sin k_y a \propto k_y$
%which is the origin of Hall effects \cite{Kontani06}.
Here, we consider the motion of a $\uparrow$-spin electron 
on the left side of Fig. \ref{fig:flux}
along a triangle of a half unit cell:
An electron in the $d_{xy}$-orbital can transfer to
the $d_{x^2\mbox{-}y^2}$-orbital and vise versa using the SO interaction
for a $\uparrow$-electron $\hbar \l {\hat l}_z/2$;
$\langle xy|{\hat l}_z|x^2\mbox{-}y^2\rangle
=-\langle x^2\mbox{-}y^2|{\hat l}_z|xy\rangle=2i$.
By considering the sign of the interorbital hopping integral 
($\pm t$ and $\pm t'$) and matrix elements of the SO interaction, 
we can verify that a clockwise (anticlockwise) motion along 
any triangle path with the SO interaction causes the factor $+i$ ($-i$).
This factor can be interpreted as the Aharonov-Bohm phase factor 
$e^{2\pi i\phi/\phi_0}$  [$\phi_0=hc/|e|$], where $\phi$ represents the 
``effective magnetic flux'' \cite{Kontani07} 
%$\varphi= (-2\pi/\phi_0)\oint{\bf A} d{\bf r}=\pm \pi/2$ [$\phi_0=hc/|e|$].
$\phi= \oint{\bf A} d{\bf r}=\pm \phi_0/4$.
This effective magnetic flux gives rise to the SHC of order $O(\lambda)$.

We also discuss the origin of the OHE by considering 
the motion of an electron with 
$|l_z=+2\rangle \propto |x^2\mbox{-}y^2 \rangle + i |xy \rangle$.
We can show that an electron with $|l_z=\pm2\rangle$ in Pt
acquires the Aharonov-Bohm phase, which gives rise to 
the OHC of order $O(\lambda^0)$ \cite{Kontani07}.

In summary, we have studied the origin of huge SHC and OHC in Pt using a
($6s,6p,5d$) tight-binding model, and found that 
the SHC reaches $1000 \ \hbar e^{-1}\cdot\Omega^{-1}{\rm cm}^{-1}$ 
in the low-resistivity regime where $\rho< 10 \ \mu\Omega{\rm cm}$.
Other significant findings of the present study are that
(i) the OHC is still larger than the SHC in Pt, which will cause
large surface magnetization of Pt;
(ii) the huge SHC and OHC originate from the effective magnetic flux 
created by the $d_{xy}$- and $d_{x^2\mbox{-}y^2}$-orbitals; and
(iii) the coherent-incoherent crossover behaviors of the SHC and OHC 
are derived by taking both the $I$-term and 
$I\!I$-term into account correctly.
When $\rho\sim100 \ \mu\Omega{\rm cm}$, 
the obtained SHC becomes comparable with the experimental value
$240 \ \hbar e^{-1} \cdot \Omega^{-1}{\rm cm}^{-1}$.
Note that $\rho$ in the present calculation corresponds to
the resistivity within the spin diffusion length ($\sim$10 nm) 
from the interface of the junction.
We comment that the effect of the overlap inregral reduces the
magnitude of the SHC for Pt to some extent \cite{comment}.
Finally, we discuss the role of the Coulomb interaction:
Although the SHC is independent of the renormalization factor 
$z=(1-\d\Sigma(\w)/\d\w)^{-1}|_{\w=0}\ (=m/m^*)$ \cite{Kontani06,Kontani07},
it will depend on the $\w$-dependence of $\g(\w)$ as well as the 
CVC due to the Coulomb interaction. 
They are important future issues.

During the preparation of this paper, we found a paper where the 
SHC was calculated based on a relativistic first-principles 
calculation \cite{Pt-naga}. 
Only the Berry curvature term given by eq. (\ref{eqn:AP-II2}) was calculated, 
which is not justified in the high-resistivity regime ($\g\gg\Delta$) 
\cite{Kontani06}.
%They did not derive our main results (i)-(iii).

The authors acknowledge fruitful discussions with Y. Otani, T. Kimura,
M. Sato, and T. Tanaka.
This work was supported by the Next Generation supercomputing Project, 
Nanoscience Program, Grant-in-Aid for the 21st Century COE 
``Frontiers of Computational Science'', and
Grant-in-Aid for Scientific Research from
the Ministry of Education, Science, Sports and Culture of Japan.

%%%%%%%%%%%%%%%%%%%%
% references
%%%%%%%%%%%%%%%%%%%%


\begin{thebibliography}{99}
\bibitem{comment}
T. Tanaka et al., arXiv:0711.1263;
in the presence of overlap integrals between different sites,
the current is
${\hat j}_\mu^C = -e\frac{\d{\hat H}_0}{\d k_\mu} 
 -e \{ {\hat H}_0, \frac{\d{\hat O}^{-1}}{\d k_\mu}{\hat O} \}/2$,
where ${\hat O}$ is the overlap integral matrix.
We verified that the second term does not change the overall
behavior of the SHC for Pt, although it reduces the magnitude of  
the SHC by approximately 50\%.
\end{thebibliography}
\end{document}